\documentclass[sigconf,nonacm]{acmart}

\AtBeginDocument{%
  }

\usepackage{multirow}
\usepackage{placeins}
\usepackage{booktabs}
\usepackage[utf8]{inputenc}
\usepackage{array}
\usepackage{booktabs}

\begin{document}

\title{Identity Emergence in the Context of Vaccine Criticism in France}

\author{Melody Sepahpour-Fard}
\orcid{}
\affiliation{%
  \institution{University of Limerick}
  \city{Castletroy}
  \state{Limerick}
  \country{Ireland}
}
\email{melody.sepahpourfard@ul.ie}

\author{Michael Quayle}
\affiliation{%
  \institution{University of Limerick}
  \city{Castletroy}
  \state{Limerick}
  \country{Ireland}}
\email{mike.quayle@ul.ie}

\author{Padraig MacCarron}
\affiliation{%
  \institution{University of Limerick}
  \city{Castletroy}
  \state{Limerick}
  \country{Ireland}
}
\email{padraig.maccarron@ul.ie}

\author{Shane Mannion}
\affiliation{%
  \institution{University of Limerick}
  \city{Castletroy}
  \state{Limerick}
  \country{Ireland}
}
\email{shane.mannion@ul.ie}

\author{Dong Nguyen}
\affiliation{%
  \institution{Utrecht University}
  \city{Utrecht}
  \state{Utrecht}
  \country{Netherlands}
}
\email{d.p.nguyen@uu.nl}


\begin{abstract}

This study investigates the emergence of collective identity among individuals critical of vaccination policies in France during the COVID-19 pandemic. As concerns grew over mandated health measures, a loose collective formed on Twitter to assert autonomy over vaccination decisions. Using analyses of pronoun usage, outgroup labeling, and tweet similarity, we examine how this identity emerged. A turning point occurred following President Macron's announcement of mandatory vaccination for health workers and the health pass, sparking substantial changes in linguistic patterns. We observed a shift from first-person singular (\textit{I}) to first-person plural (\textit{we}) pronouns, alongside an increased focus on vaccinated individuals as a central outgroup, in addition to authority figures. This shift in language patterns was further reflected in the behavior of new users. An analysis of incoming users revealed that a core group of frequent posters played a crucial role in fostering cohesion and shaping norms. New users who joined during the week of Macron's announcement and continued posting afterward showed an increased similarity with the language of the core group, contributing to the crystallization of the emerging collective identity.

\end{abstract}




\maketitle

\section{Introduction}
Group identity, i.e., the part of our identity tied to group membership, is a fundamental aspect of human experience that shapes our emotions and behaviors~\cite{BaumeisterLeary1995The}. It facilitates the motivation and collective power to act as a group~\cite{BaumeisterLeary1995The, brewer1991social}, making it central to collective action and social movements~\cite{taylor1992collective, reicher1996battle, simon1998collective}. These socially negotiated collective identities influence and reshape our societies. Therefore, understanding the emergence of collective identities is essential to inform policy makers and activists about the mechanisms driving these identities and their potential societal impacts.

\begin{figure}
    \centering
    \includegraphics[width=1\columnwidth]{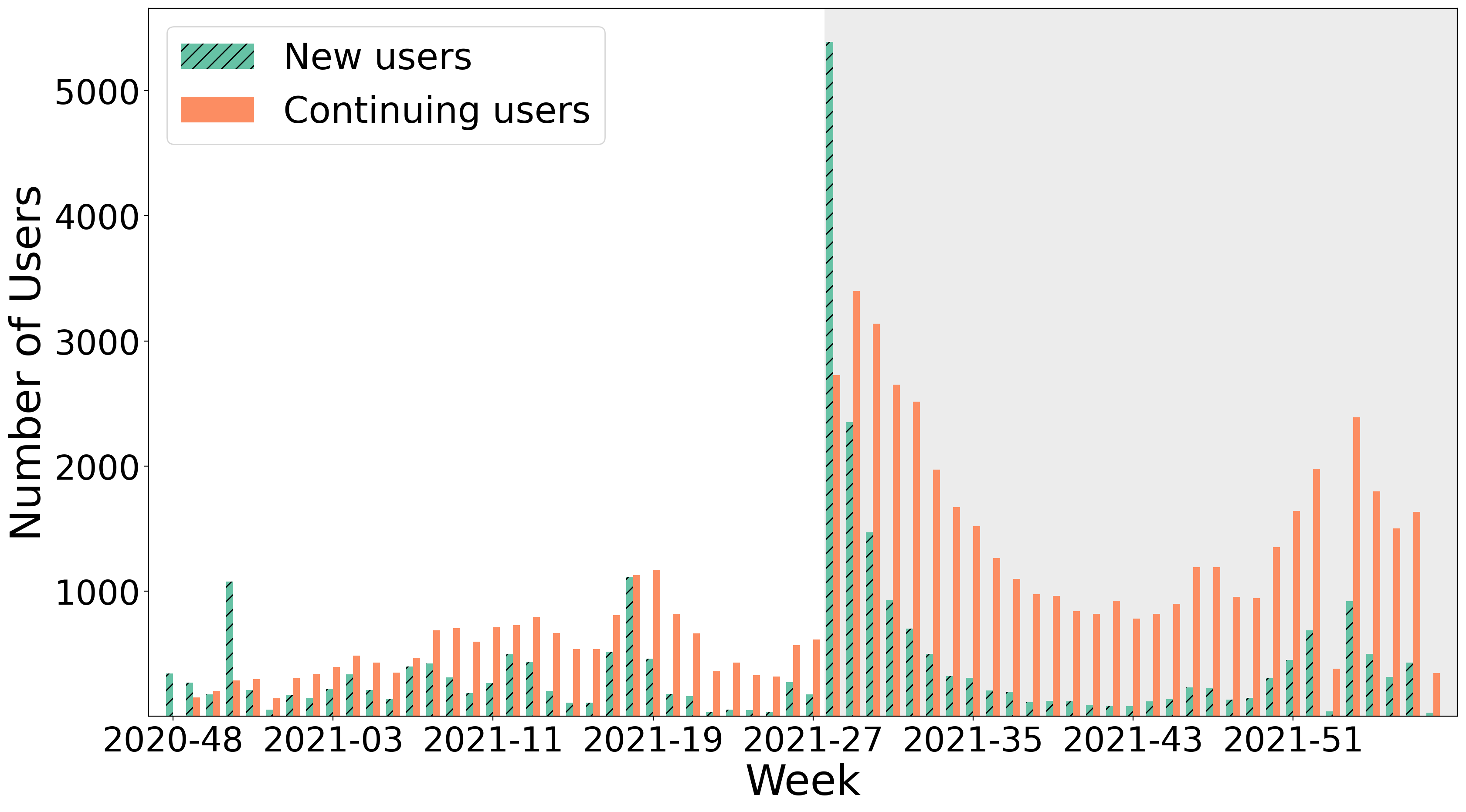}
  \caption{Number of new and continuing vaccine-critical users by weeks, where a new user is posting for the first time and a continuing user has posted at least once before. The grey-shaded area represents the period following Macron's announcement on mandatory vaccination for health workers and the health pass, which coincided with a surge in new users joining the conversation.}
  \Description{The figure represents the number of new and continuing users each week from the end of 2020 to the the beginning of 2022. A new user posts for the first time and a continuing user has posted at least once before. We observe a relatively stable number of new users on the first half of the time frame until Macron's speech on July 12th, 2021. That week, we see an important surge of new users and the number of total users posting stays high until the end of the time frame.}
  \label{teaser}
\end{figure}

Social media platforms like Twitter (now X) provide a unique window into the dynamics of groups and social movements, offering vast amounts of naturalistic data that capture real-time interactions among users. Previous research has leveraged social media to explore a variety of social phenomena, including political participation~\cite{gil2014social, theocharis2023platform}, polarization~\cite{kubin2021role, tucker2018social}, social influence~\cite{gonzalez2020social}, and self-presentation~\cite{sepahpour2022mothers, sepahpour2023using}. Furthermore, social media enables researchers to observe how individuals express their identities, form ingroups and outgroups, and respond to significant events such as the COVID-19 pandemic and related policy changes. 

While numerous studies have leveraged social media to explore the impact of COVID-19, particularly in terms of social media influence~\cite{gonzalez2020social, venegas2020positive, hussain2020role} and misinformation spread~\cite{hossain2020covidlies, vraga2021addressing, al2021covid}, relatively few have examined the anti-vaccine identity~\cite{motta2023identifying, KADICMAGLAJLIC2024116721}. Moreover, to our knowledge, none have specifically focused on the \textit{process} of collective-identity emergence among vaccine-critical individuals. 

The high level of vaccine skepticism makes France a particularly relevant context for the study of identity emergence linked to vaccine criticism. The Wellcome Global Health Monitor identified France as one of the most vaccine-skeptical countries in the world~\cite{wellcome}. This skepticism is deeply rooted in historical distrust of the pharmaceutical industry, past vaccine controversies, and a strong cultural emphasis on individual freedom~\cite{ward2022french}. The introduction of the health pass and mandatory vaccination for health workers by President Emmanuel Macron on July 12, 2021, intensified these sentiments, with many viewing it as an overreach of government authority and an infringement of personal freedom~\cite{faccin2022assessing}. Notably, even among vaccinated individuals, doubts about the vaccine increased from 44\% to 61\% following the health pass implementation~\cite{ward2022french}. These sentiments have fueled widespread protests~\cite{LeMonde2021} and led to the formation of online communities dedicated to opposing vaccine mandates and promoting skepticism~\cite{peretti2020future}. Only a few studies have analyzed the content of French anti-vaccine tweets and the users they attract~\cite{faccin2022assessing}, the impact of mandatory vaccination on vaccination rates, and anti-vaccine arguments~\cite{sauvayre2023obligation, gable2023fight}. 

This gap in the literature on the study of identity emergence may stem from the challenges associated with studying such a fluid and complex phenomenon~\cite{morselli2022digital}. Computational methods and large-scale social media data offer a valuable opportunity to explore the dynamic emergence of identity and its underlying mechanisms.

\textbf{Present Work.} We explore the emergence and evolution of collective identity among Twitter users critical of COVID-19 vaccines and related policies in France, highlighting the role of significant events in catalyzing shifts in group identity and the importance of language in these processes. 

\textbf{Data and Methods.} 
We analyze a dataset of 338,641 tweets posted between December 2020 and January 2022 that contains specific hashtags associated with vaccine criticism in France. First, we examine pronoun usage by measuring the prevalence of first-person \textit{singular} (e.g. \textit{je}/\textit{I}) and \textit{plural} (e.g., \textit{nous}/\textit{we}) pronouns in tweets. Then, we identify outgroup labels based on their cosine similarity to third-person plural pronouns (i.e., \textit{they} and \textit{them}) in word embedding models~\cite{mikolov2013distributed}. Finally, we use cosine similarity of sentence embeddings~\cite{martin-etal-2020-camembert} and the Fightin' Words method~\cite{monroe2008fightin} to investigate the language of different groups of users.

\textbf{Results.} Overall, our findings revealed that President Macron's speech, on July 12th, 2021, acted as a catalyst, leading to the consolidation of collective identity. Figure \ref{teaser} shows the surge in the number of users criticizing COVID-19 vaccines and related policies following his speech. 
We identified several key components of collective identity emergence:

\textbf{(1) Pronoun use}: Before Macron's announcement, tweets predominantly used first-person singular pronouns, reflecting individual perspectives. In contrast, following the announcement, there was a marked shift towards first-person plural pronouns, indicating an emerging sense of collective identity.

\textbf{(2) Outgroups}: Our analysis showed that the users' primary focus was on authority figures. Over time, we also observed an increasing focus on vaccinated individuals. 

\textbf{(3) Incoming members and Conformism}: Users who joined the discussion during the week of Macron's speech—when the influx of new users was highest (see Figure \ref{teaser})—and continued posting afterward exhibited increased similarity with established users. This trend may indicate a tendency to conform to the group’s linguistic norms and assert their identity within the movement~\cite{nguyen-p-rose-2011-language, 10.1145/2488388.2488416}.

\section{Related Work}

We contextualize our analysis by reviewing key studies and theories on collective identity. First, we define collective identity (\S\ref{definition}), then examine the shift from personal to collective identity (\S\ref{shift identity}) and the role of outgroups (\S\ref{outgroupslit}). Finally, we discuss research on the dynamics of new members joining established groups (\S\ref{newentrants}).

\subsection{Collective Identity: Definition} \label{definition}

Definitions of collective identity vary across the literature. \citet{polletta2001collective} define it as "the individual's cognitive, moral, and emotional connection with a broader community, category, practice, or institution" (p.285).\footnote{This definition, with its focus on the individual’s perspective, also aligns with the concept of social identity as defined by social psychologists~\cite{tajfel1979integrative}. In this study, we draw from both the Sociology and Social Psychology literature to explore group identity processes.} Other scholars, such as \citet{snow2001collective}, view collective identity as an interindividual process, whereby it emerges through interactions and actions~\cite{snow2001collective, whittier1995feminist}, as well as through shared interests, experiences, and solidarity, leading to a shared definition~\cite{taylor1992collective} and a sense of belonging to a group~\cite{smithey2009social}. 
The identification process can lead to self-stereotyping, in which individuals amplify their similarities with the ingroup, adopt prototypical behaviors, and conform to the ingroup norms~\cite{simon1994self, tajfel1979integrative, moreland1985social} (e.g., through common language~\cite{flesher2010collective}). 

In relation to social movements, collective identity helps conceptualize how individuals unite, coordinate, and commit within a movement, as well as how these movements emerge and persist~\cite{flesher2010collective}. It is regarded as a crucial element for the cohesion and success of social movements~\cite{melucci1980new, melucci1985symbolic, melucci1988getting, melucci1996challenging, melucci2013process}. Previous research has explored collective identity on social media in various movements, including the Yellow Vests movement in France~\cite{luders2022becoming, morselli2022digital}, the Iranian 'My Stealthy Freedom' movement~\cite{khazraee2018digitally}, the responses to Russia’s invasion of Ukraine~\cite{o2024strategic}, and the Stop the Steal campaign following the 2020 US presidential election~\cite{spann2023evaluating}. These studies have employed both qualitative methods~\cite{khazraee2018digitally} and quantitative approaches such as topic modeling~\cite{morselli2022digital, spann2023evaluating} and network analysis~\cite{spann2023evaluating}. For example, \citet{spann2023evaluating} analyzed the Stop the Steal campaign on Twitter by using topic modeling to track the evolution and alignment of discussion topics, and social network analysis to evaluate the structural cohesion of the community over time. Additionally, despite some exceptions (e.g.,~\cite{rousseau2005emergence, drury2000collective}), empirical studies, particularly in Social Psychology, have often treated collective identity as a predefined construct guiding collective action~\cite{van2012conviction, van2018integrating}. In contrast, the present study uses multiple linguistic indicators—such as pronoun use, outgroup labeling, tweet similarity, and distinctive words—alongside large-scale social media data to capture the dynamic and evolving nature of collective identity through language.

\subsection{From a Personal to a Collective Identity} \label{shift identity}

The emergence of collective identity marks a shift from a personal (i.e., "I") to a collective (i.e., "we") locus of self-definition~\cite{brewer1996we, taylor1986two}. The choice between first-person singular and plural pronouns reflects individuals' relationship with their audience~\cite{maitland1987pronominal}, showing how language can be a central part of identity formation in interaction~\cite{labov2011principles}. The first-person plural pronoun "we", in particular, conveys a sense of inclusion and belonging, activating and emphasizing shared identity among speakers~\cite{pennebaker2011secret, brewer1996we, papapavlou2009relational}. 
   
Several studies have examined the relationship between first-person plural pronouns and group-identity orientations~\cite{pennebaker2013counting, rafaeli1997networked, michinov2004social, lee2020fandom, nguyen-p-rose-2011-language}. For instance, \citet{smith2015social} found that the increased use of the first-person plural pronoun "we" was associated with agreement on the norms in the Occupy Wall Street movement. In contrast, the more people contribute to a community and align with its norms, the less they use first-person singular pronouns~\cite{10.1145/2488388.2488416}. Furthermore, \citet{lee2020fandom} examined group identity within a Kpop fandom on Facebook and observed that the increased use of "we" over "I" for self-referencing not only predicted higher levels of group interaction, evidenced by increased engagement such as likes and comments, but also facilitated the consolidation of a cohesive group identity by fostering a shared sense of belonging and collective action within the fandom.

\subsection{Outgroups and Comparison} \label{outgroupslit}

When defining one's identity, two criteria of comparison are involved: sameness and distinctiveness~\cite{van2013collective, pickett2006using}. While sameness can be achieved within the ingroup, distinctiveness is achieved through comparison with relevant outgroups~\cite{brewer1991social, pickett2006using}. 
When in conflict with an outgroup, the perceived opposition makes groups tend towards ideas and actions in reaction to each other~\cite{van2013collective}. For instance, \citet{drury2000collective} found that confrontations with an outgroup, such as the police during protests, could unite otherwise divided protest participants and radicalize them. Research has also highlighted how institutionalized rejection of a group based on identity can spark major social movements~\cite{marx1998making}.

Relative deprivation, the perception of being unfairly disadvantaged compared to others, is another factor influencing social movements, intergroup attitudes, and collective action~\cite{luders2021bottom, guimond2002prosperity, abrams2012testing}. The theory of relative deprivation states that people do not make absolute judgment on fairness but rather perceive a situation as fair or unfair in comparison to outgroups~\cite{crosby1976model}. This comparison can be vertical, with a group having a position of power (e.g., the "elite'' or an authority), or horizontal, with groups of comparable status (e.g., vaccinated people for non-vaccinated ones)~\cite{luders2021bottom}. Research has shown relative group deprivation, both vertical and horizontal, to be positively related to protest participation~\cite{luders2021bottom}.

\subsection{Established and Incoming Users} \label{newentrants}
In a social movement, incoming members may adopt the behaviors and language of established members to enhance their sense of belonging to the group. When individuals enter a group, they are met with certain expectations that may influence their attitudes and behaviors, regulating their assimilation to the group~\cite{moreland1985social}. Established ingroup members will define the norms and therefore how to think and act for incoming group members~\cite{livingstone2011we, levine2010encyclopedia}.

Additionally, incoming members’ motivation to join the group can vary. High identifiers, who are strongly motivated to be part of the group, are more likely to exert extra effort to be accepted as legitimate members~\cite{branscombe1999context}. They will often engage in prototypical behaviors and align their language with ingroup norms to gain favorable recognition~\cite{noel1995peripheral}, while using derogatory language to differentiate themselves from the outgroup~\cite{branscombe1999context, turner1991social, potter1987discourse}. In contrast, low identifiers, who are less invested in the group, may exhibit less alignment~\cite{branscombe1999context}. For instance, \citet{10.1145/2488388.2488416} investigated how incoming users integrate into communities with established norms and found that linguistic conformity with ingroup norms predicts incoming users' commitment to the group.

\section{Dataset}
To explore the emergence of collective identity, we focus on individuals criticizing COVID-19 vaccines and related policies in France during the pandemic. In this section, we describe our Twitter data collection (\S\ref{datacol}) and data preparation steps (\S\ref{dataprep}). Although we report this analysis and give examples in English, raw data and the analysis pipeline were in French.

\subsection{Data Collection} \label{datacol}
Based on previous research that widely uses hashtags for Twitter data collection~\cite{reavley2014use, yardi2010tweeting, linabary2020feminist, o2024strategic}, and given that hashtags enable a more precise targeting of specific topics and audiences~\cite{conover2011political}, we choose to base our data collection on hashtags. We compile a list of hashtags associated with vaccine criticism through a qualitative search on Twitter (i.e., by searching known hashtags and manually gathering co-occurring hashtags from tweets seemingly posted by vaccine-critical users), as well as from various websites and blog posts.\footnote{The blogpost "Les mouvements anti pass et antivax sur les réseaux sociaux en France: les hashtags" ("The anti-pass and antivax movements on Social Media in France: the hashtags") written by Christophe Asselin's and posted on January 25, 2022, provided a set of hashtags used by French antivax and anti-pass users that we included in our list. Internet archive: http://web.archive.org/web/20230209102632/https://blog.digimind.com/fr/agences/
mouvements-anti-pass-antivax-reseaux-sociaux-france-etude-hashtags} The finalized set contains the following hashtags: \#PassSanitaire (health pass), \#VaccinationObligatoire (mandatory vaccination), \#antivax, \#antivaccin, \#antivaxx, \#NonAuVaccin (no to vaccine), \#JeNeMeFaisPasVacciner (I am not getting vaccinated), \#JeNeMeFeraisPasVacciner (I won't get vaccinated), \#PassDeLaHonte (shame pass), \#NonAuPassDeLaHonte (no to the shame pass), \#NonAuPassSanitaire (no to the health pass), \#NonAuPassVaccinal (no to the vaccination pass), \#DictatureSanitaire (health dictatorship), \#StopDictatureSanitaire (stop health dictatorship), \#NousSommesDesMillions (we are millions). 

Using the Twitter API v2 for academic research, we collect all tweets and retweets containing at least one of the hashtags from December 2020 to January 2022. Our dataset encompasses a total of 2,177,538 posts. In our analysis, we specifically focus on original tweets and exclude retweets (i.e., posts starting with "RT"), because we want to analyse the active participation in the creation of linguistic content. This results in a dataset of 402,836 tweets authored by 38,998 unique users.

\subsection{Data Preparation} \label{dataprep}
We clean the data by eliminating mentions, URLs, duplicated punctuation, extra spaces, and tweets consisting solely of punctuation, and we remove duplicates. Additionally, we remove tweets censored by Twitter for Terms of Service violation since their content was deleted (and replaced by the text "[User] account is temporarily unavailable because it violates the Twitter Media Policy. Learn more."), resulting in a refined dataset comprising 401,067 tweets authored by 38,836 unique users.

Following a manual review of the dataset, we notice that the collected dataset contains tweets from pro-vaccination users who promote vaccination and criticize vaccine-hesitant users. As we study the emergence of collective identity within the group formed by people criticizing COVID-19 vaccines and related policies, we need to remove tweets posted by other groups from our dataset. We identify tweets that do not align with our study focus, including: a) pro-vaccination tweets explicitly advocating for the vaccine or criticizing vaccine hesitancy; b) tweets seemingly not representing individual users; c) tweets criticizing both pro- and anti-vaccine positions; d) unrelated tweets, i.e., tweets using at least one of the hashtag used for data collection but with a content unrelated to COVID-19 or vaccination. To remove those tweets, we train a classifier on manually classified tweets (1\% of the total dataset). Given that our data is in French, we fine-tune CamemBERT for sequence classification~\cite{martin-etal-2020-camembert}, which is based on the RoBERTa architecture~\cite{liu2019roberta} and specifically trained and evaluated on French data. The classifier achieved a precision of .97, a recall of .96, and an F1-score of .97 for predicting the class of interest: tweets related to the criticism of COVID-19 vaccines and related policies. We use our classifier on the remaining unlabeled dataset and filter out non-relevant tweets (15.56\% of tweets). Our final dataset for all subsequent analyses consists of 338,641 tweets and 27,016 users (see Appendix for more details on classification).

\section{From \textit{I} to \textit{we}} \label{pronounsection}

Pronoun use, particularly first-person singular and plural pronouns, reveals the authors' relationships with their audience (see \S\ref{shift identity}). We examine the prevalence of these pronouns in our data and how they reflect the emergence of collective identity.

\subsection{Methods}

We compile a comprehensive list of pronouns categorized by the person to whom they refer: first-person singular (1SG, e.g., \textit{je}/\textit{I})\footnote{Full list of first-person singular pronouns: \textit{je} (I), \textit{j'} (I), \textit{me} (me), \textit{moi} (me), \textit{mien} (mine), \textit{mienne} (mine), \textit{miens} (mine), \textit{miennes} (mine), \textit{ma} (my), \textit{mon} (my)} and first-person plural (1PL, e.g., \textit{nous}/\textit{we})\footnote{Full list of first-person plural pronouns: \textit{nous} (we/us), \textit{notre} (our), \textit{nos} (our), \textit{nôtre} (ours), \textit{nôtres} (ours)}. Subject (e.g., \textit{I}), object (e.g \textit{me}), and possessive (e.g., \textit{mine}) pronouns are included in our analysis. To analyze the trajectory of pronoun usage, we identify occurrences of pronoun groups based on two criteria: a) if any pronoun (subject, object, or possessive) appears within the body of a tweet, excluding hashtags, and b) if a subject pronoun (i.e., \textit{I} or \textit{we}) is at the beginning of a hashtag. If either of these conditions is met, the tweet is labeled as 1 (present) for that pronoun group; otherwise, it is labeled as 0 (absent). A single tweet can contain both pronoun groups. We then use proportion z-tests to compare pronoun use before and after Macron’s July 12, 2021, announcement on mandatory vaccination and the health pass.

\subsection{Results}

\begin{figure*}[!htbp]
    \centering
      \begin{minipage}{0.485\linewidth}
          \centering
          \includegraphics[width=\linewidth]{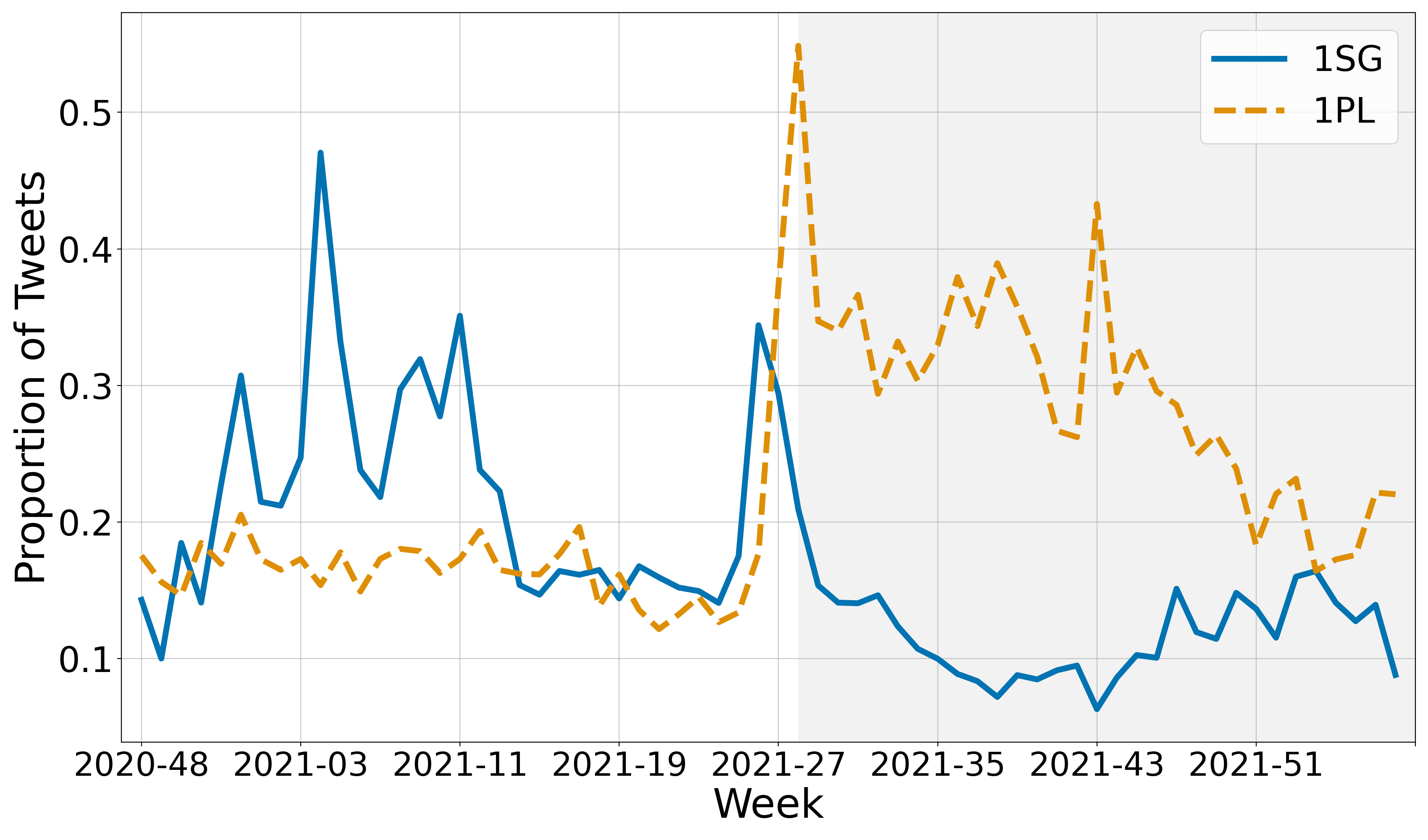}
          \textbf{A}
      \end{minipage}%
      \hspace{0.3cm}
      \begin{minipage}{0.485\linewidth}
          \centering
          \includegraphics[width=\linewidth]{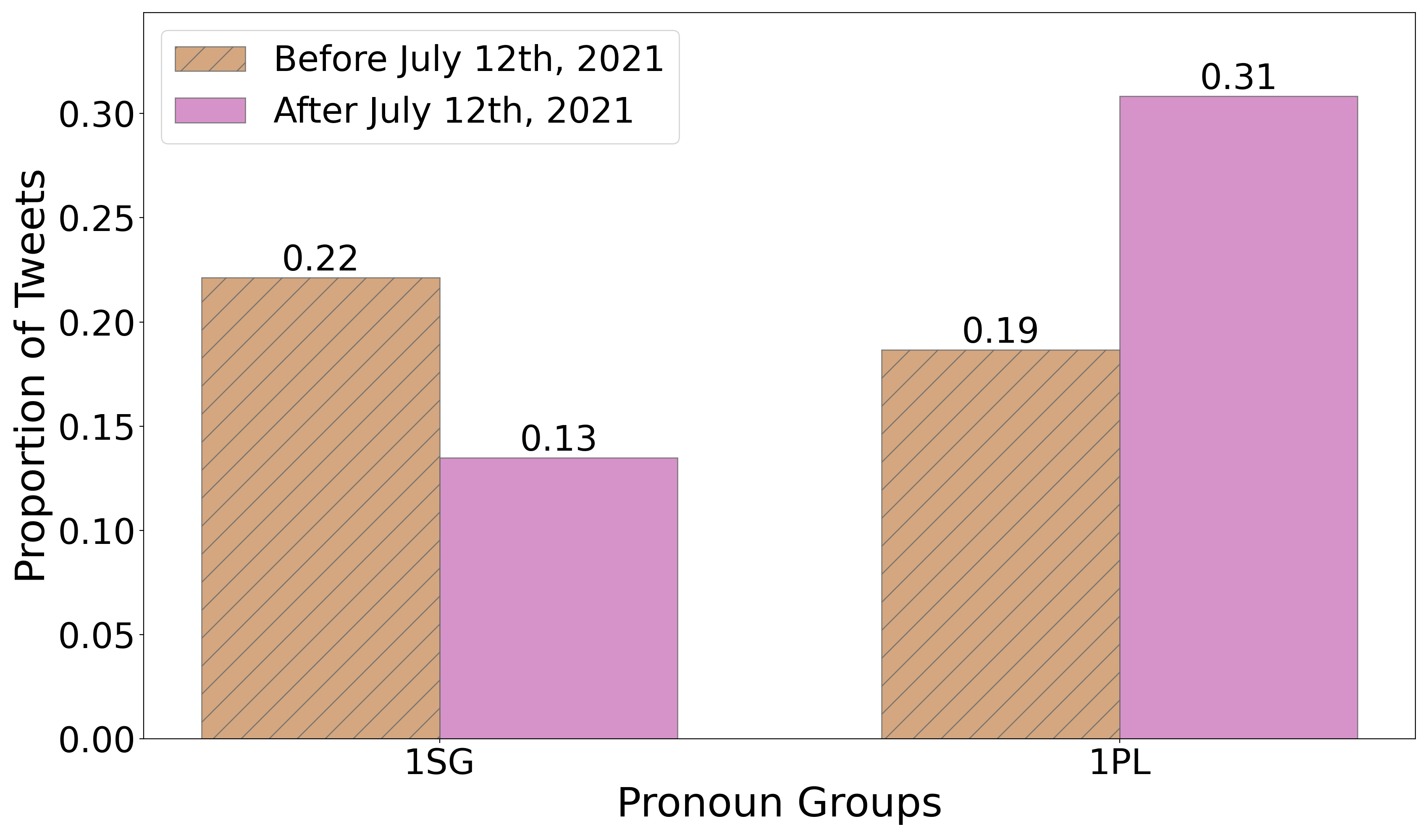}
          \textbf{B}
      \end{minipage}%
    \caption{Pronoun use (A) across weeks, with the period following Macron's speech shaded grey, and (B) before versus after Macron’s speech, considering two groups of pronouns: first-person singular (1SG) and first-person plural (1PL) pronouns. Before Macron’s speech, users mostly used 1SG pronouns while they used more 1PL pronouns after.}
    \label{pronounuse}
    \Description{These figures illustrate the prevalence of first-person singular and plural pronouns in tweets. Figure A tracks the weekly usage of these pronouns from late 2020 to early 2022, highlighting changes over time. Figure B provides a direct comparison between the proportion of tweets using these pronouns before and after Macron’s July 12, 2021, announcement on mandatory vaccination and the health pass. Prior to Macron’s speech, first-person singular pronouns were more prevalent, whereas after the speech, there is a noticeable shift, with first-person plural pronouns becoming more dominant.}
\end{figure*}

Figure \ref{pronounuse} shows the evolution of the use of pronouns (A) across weeks and (B) before versus after Macron's speech. Initially, users mostly used first-person singular pronouns such as "I" or "my". The major shift in pronoun use happened the week before Macron's speech (2021-27), with a peak on the week of the speech (2021-28). First-person plural pronouns such as "we" or "us" became the most used pronouns and they were constantly the most used pronoun group until the end of the time frame.

Figure \ref{pronounuse}B shows a important shift in pronoun use before and after Macron's speech. Before the speech, 1SG pronouns were used in 22\% of tweets, dropping to 13\% afterward, while 1PL pronouns increased from 19\% to 31\%. These changes were statistically significant, with 1SG pronouns showing a decrease (Z=60.45, \textit{p}=0.0) and 1PL pronouns showing an increase (Z=-69.49, \textit{p}=0.0). 

We conducted supplementary analyses, grouping tweets by authors, removing hashtags from the body of tweets (i.e., tokens starting with \textit{\#}), solely examining hashtags, and excluding hashtags used in data collection to check for the robustness of the analysis and the influence of hashtags. We found hashtags to have a crucial role in pronoun prevalence. If hashtags were removed from tweets, the trends did not maintain. In other words, it was mainly through hashtags that we observed the shift in pronoun use. To ensure the trend were not an artefact of our hashtag selection, we removed the hashtags used for data collection and kept the other hashtags, and the results persisted, i.e., we observed the shift from 1SG pronouns to 1PL pronouns (see Appendix for more detail on these analyses). 

The importance of 1PL ("we") pronouns in constructing collective identity is well-documented~\cite{maitland1987pronominal, pennebaker2011secret, lee2020fandom, rafaeli1997networked, michinov2004social, papapavlou2009relational, smith2015social}. The shift from 1SG to 1PL pronouns in our findings suggests the emergence of a collective identity among French-speaking Twitter users critical of COVID-19 vaccines and related policies. Before Macron's speech, users primarily expressed individual perspectives through 1SG pronouns. However, the July 12, 2021, announcement appears to have triggered a shift towards defining oneself in relation to the ingroup~\cite{turner1994self}, as reflected in the increased use of 1PL pronouns.

\section{Who are \textit{they}?} \label{outgroupsection}

We track the prevalence of outgroup references and analyze the types of outgroups mentioned over time. We then explore how these patterns relate to the formation of collective identity.

\subsection{Methods}

To complement the analysis of how individuals identify with their ingroup through the use of pronouns, we now focus on how they identify and define their outgroups through labels. To analyze how ingroup members approach and frame their outgroup, \citet{yoder-etal-2023-identity} studied a online community and gathered identity terms referring to minority groups from different sources and extended the list using word embeddings. Inspired by the latter, we develop a method based on word embeddings and qualitative analysis of tweets to identify outgroup labels. We first preprocess our data for creating word embeddings. Our preprocessing involves removing nonalphabetic characters (except for hashtags) and converting tweets to lowercase. Furthermore, we observe frequent occurrences of the phrase "non vacciné" (non-vaccinated) wherein "non" precedes "vacciné" to negate it. To treat this phrase as a distinct token, we insert an underscore between the two words (i.e., "non\_vacciné").\footnote{We use a similar process with related words such as injected, vax, spiked, dosed, etc.}

We train two distinct word embedding models using Word2Vec~\cite{mikolov2013distributed}: one for tweets posted before July 12, 2021, and another for those posted from July 12, 2021, onward. This division is motivated by the sharp increase in tweet volume after this date (see Figure~\ref{teaser}).\footnote{Had we created a single model for the entire period, the significant post-event data could have skewed the embeddings, potentially obscuring outgroup labeling practices that existed prior to the event.} We then examine words most similar to the pronouns "ils" (they) and "eux" (them) using cosine similarity and which appear at least 20 times to initiate a qualitative selection of outgroup labels. Expanding on this initial label selection, we examine tokens most similar to those identified in the first round, resulting in 126 outgroup labels. Finally, to ensure the selected outgroup labels accurately correspond to the outgroup, we sample 20 tweets for each label found and deem it suitable for selection if at least 75\% of tweets using the label, did so to refer to the outgroup. 

\subsection{Results}

We found 98 outgroup labels in total that we manually classified into six categories depending on the entity they referred to (i.e., authority, people, media) and their stance (i.e., neutral or critical): Authority - Neutral (n=13), Authority - Critical (n=43), People - Neutral (n=6), People - Critical (n=31), Media - Neutral (n=2), Media - Critical (n=3). The complete list of categorized outgroup labels can be found in the Appendix. Table \ref{outgroup_table} shows examples of outgroup labels classified into the different categories.

\begin{table*}[!ht]
    \centering
    \small
    \begin{tabular}{@{}c p{5.3cm} p{5.3cm} p{4.7cm}@{}}
        \toprule
        & \textbf{Authority} & \textbf{People} & \textbf{Media} \\
        \midrule
        \multirow{2}{*}{\textbf{Neutral}} 
        & gouvernement (government), élites (elites), dirigeants (leaders), autorités (authorities), flics (cops), laboratoires (laboratories), président (president), ministres (ministers), leaders, labos (labs), etc.
        & vaccinés (vaccinated), vaxxinés (vaxxinated), spikés (spiked), dosés (dosed), vaxxés (vaxxed), provax
        & médias (media), journalistes (journalists) \\
        \cmidrule{2-4}
        \multirow{2}{*}{\textbf{Critical}} 
        & dictateurs (dictators), tyrans (tyrants), nazis, collabos (collaborators with the Nazi regime),  macronistes (Macron supporters), macronards (derogatory term for Macron supporters), talibans, ayatollahs, khmers, manipulateurs (manipulators), etc.
        & endoctrinés (indoctrinated), moutons (sheep), covidistes (covidists), lobotomisés (lobotomized), lâches (cowards), mougeons (blend of sheeps and morons), terrorisés (terrorized), ignorants (ignorant), incultes (uncultured), obéissants (obedient),  etc.
        & merdias (derogatory term for media), journaleux (biased journalists), journalopes (derogatory term for journalists) \\
        \bottomrule
    \end{tabular}
    \captionof{table}{Examples of outgroup labels by category. Outgroup labels were manually categorized through two dimensions: the entity they refer to (authority, people, or media) and their stance about the entity (neutral or critical).}
    \label{outgroup_table}
\end{table*}

A clear pattern emerged in the disparity between neutral and critical labels. While there were relatively few neutral ways to refer to authority or the people, there was a much larger variety of neologisms and critical expressions used to describe the outgroup. For example, critics of COVID-19 policies often targeted the broader perceived dictatorial context (e.g., "dictateurs" [dictators], "tyrans" [tyrants]), focusing specifically on Macron (e.g., "macronistes" [Macron supporters], "macronards" [derogatory term for Macron supporters]) or invoking historical conflicts (e.g., "nazis", "collabos" [collaborators with the Nazi regime], "khmers") and controversial leaders (e.g., "ayatollahs", "talibans"). When referring to vaccinated people, users frequently portrayed them as indoctrinated or lacking autonomy and courage (e.g., "endoctrinés" [indoctrinated], "lobotomisés" [lobotomized], "terrorisés" [terrorized], "obéissants" [obedient], "moutons" [sheeps]). Finally, only a few labels referred to the media, with critical ones typically being neologisms that merged a neutral term with a derogatory one (e.g., the suffix "-eux" in "journaleux" adds a negative connotation resulting in a word meaning "biased journalists").

\begin{figure}
    \centering
    \includegraphics[width=0.99\columnwidth]{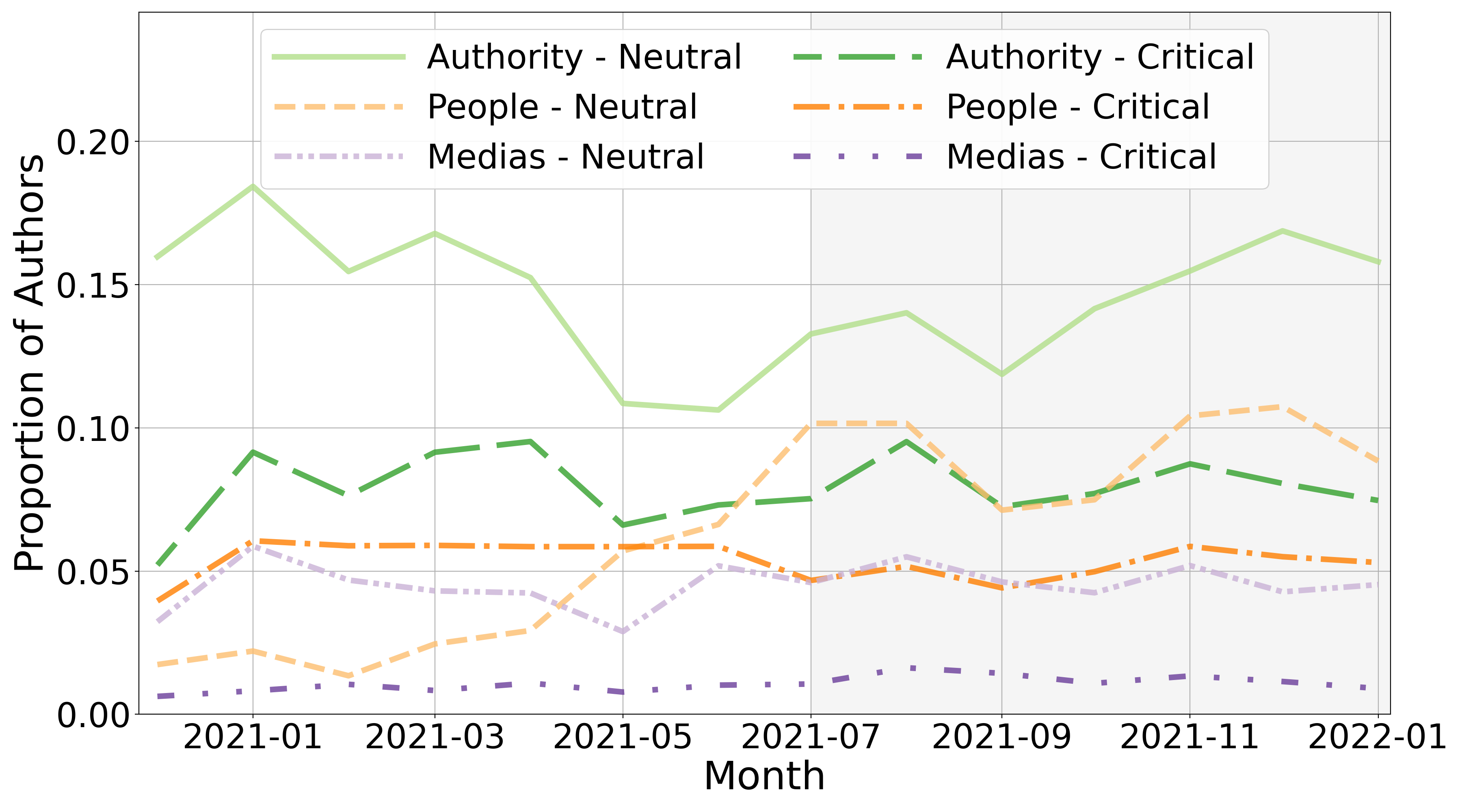}
    \caption{Proportion of authors using outgroup labels by category across months, with the period following Macron's speech shaded grey. Labels related to authority are used by more authors and the focus on vaccinated people increases over time.}
    \label{outgroup_evolution}
    \Description{The figure shows the evolution of outgroup labels by category over time. "Authority - Neutral" consistently remains the most used category. In the first half of the time frame (before Macron's speech on July 12, 2021), "Authority - Critical" is the second most used category, followed by "People - Critical" in third place. Over time, the prevalence of outgroup labels neutrally referring to vaccinated individuals (i.e., "People - Neutral") increases, becoming the second most prominent category from July 2021 onward. Labels neutrally referring to the media (i.e., "Media - Neutral") are used at similar rates to "People - Critical," while "Media - Critical" labels are only marginally used throughout the period. }
\end{figure}

Figure \ref{outgroup_evolution} shows the evolution of the different categories of outgroup labels over months in authors' tweets. Overall, the outgroup focus was mostly on authority. The outgroup labels neutrally referring to authority were the most used labels over the whole time frame. During the first half of the time frame, outgroup labels criticizing authority constituted the second most used category. Interestingly, while most proportions remained stable across categories, we observed a gradual increase in neutral outgroup labels referring to vaccinated individuals. On July 2021, the latter became the second most used category of outgroup labels. Finally, outgroup labels related to the media were the least used and quite stably. 

The study of outgroup labels allowed us to understand how ingroup members define and discuss their outgroups over time. We found that vertical references to those in power, such as the government, elected officials, and the police, were a primary focus of the group we studied. Although \citet{luders2021bottom} did not find significant differences between horizontal and vertical outgroup comparisons in the context of the French Yellow Vests movement, our results indicated that vertical comparisons were more prevalent in this context. In the case of COVID-19 vaccination, the government's decisions directly impacted them and their perception of freedom and bodily autonomy. The importance of authority also explains why Macron's speech, imposing limitations on freedom for the unvaccinated, was such an impactful event shaping the emergence of collective identity. Moreover, we observed an increasing focus on the vaccinated group as an outgroup, particularly following Macron's speech. The centrality of the authority and vaccinated outgroups can be interpreted through the lens of relative deprivation theory~\cite{crosby1976model}. Individuals critical of COVID-19 vaccines and related policies might have felt a growing sense of unfairness when it comes to the way their ingroup was treated by the authority compared to the vaccinated outgroup, which might be seen as the privileged group. Macron speech might have acted as an extreme example of such felt unfairness.

\section{Incoming Users and Conformism} \label{incomingsection}

We focus on the week of Macron's address on mandatory vaccination and the health pass, which triggered an influx of new users (see Figure \ref{teaser}) to examine how users initiate interactions and adapt to community norms.

\subsection{Methods}

To analyze the linguistic behavior of incoming users and compare it to established users, we group users based on when they began posting and their posting frequency. From the week before Macron's speech to the week after, we categorize users into four distinct groups:

\begin{itemize}
  \item Established-Core (n=427): The top 10\% of users who posted the most (out of all users in the dataset) and were active for at least ten different months, thus having posted before Macron's speech;
  \item Established-Irregular (n=8931): Users who started posting before Macron's speech but not as often as the Established-Core;
  \item Incoming-Persistent (n=2068): Users who started posting on the week of Macron's speech and continued after;
  \item Incoming-Transient (n=3321): Users who started posting on the week of Macron's speech but did not post after.
\end{itemize}

We use a sentence embedding model pre-trained on French data (Sentence-CamemBERT-base,~\cite{dangvantuan2023sentencecamembert}) and based on CamemBERT~\cite{martin-etal-2020-camembert} and Sentences-transformers~\cite{reimers2019sentence} to create sentence-embeddings for each tweet. The sentence embeddings are used to calculate the cosine similarity between tweets within and across user groups. This approach enables us to analyze how closely aligned the language of incoming users is to the one of established ones.

To better understand the distinctiveness of each group, we complement our analysis with Fightin' Words~\cite{monroe2008fightin}, implemented through the Convokit package~\cite{chang2020convokit}. It uses Bayesian techniques to adjust the estimated differences in word usage between groups, ensuring more reliable comparisons by pulling extreme values toward more reasonable estimates (shrinkage) and preventing the model from being influenced by rare or noisy data (regularization).

\subsection{Results}
\begin{figure*}[!ht]
  \centering
  \raggedbottom  
  \begin{minipage}[b]{0.235\linewidth}
      \centering
      \includegraphics[width=\linewidth]{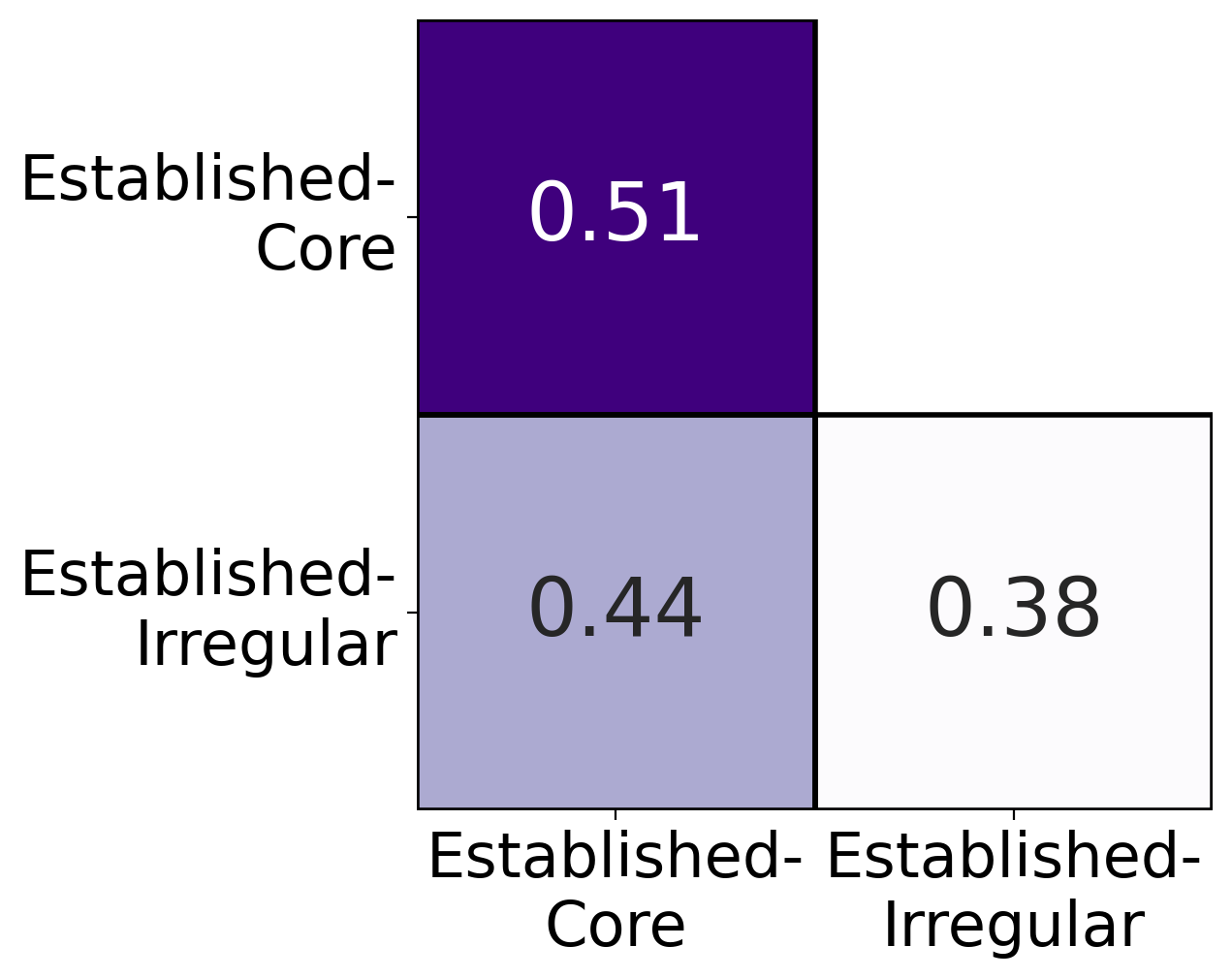}
      \textbf{A - Week 2021-27}
  \end{minipage}%
  \hspace{0.2cm}
  \begin{minipage}[b]{0.39\linewidth}
      \centering
      \includegraphics[width=\linewidth]{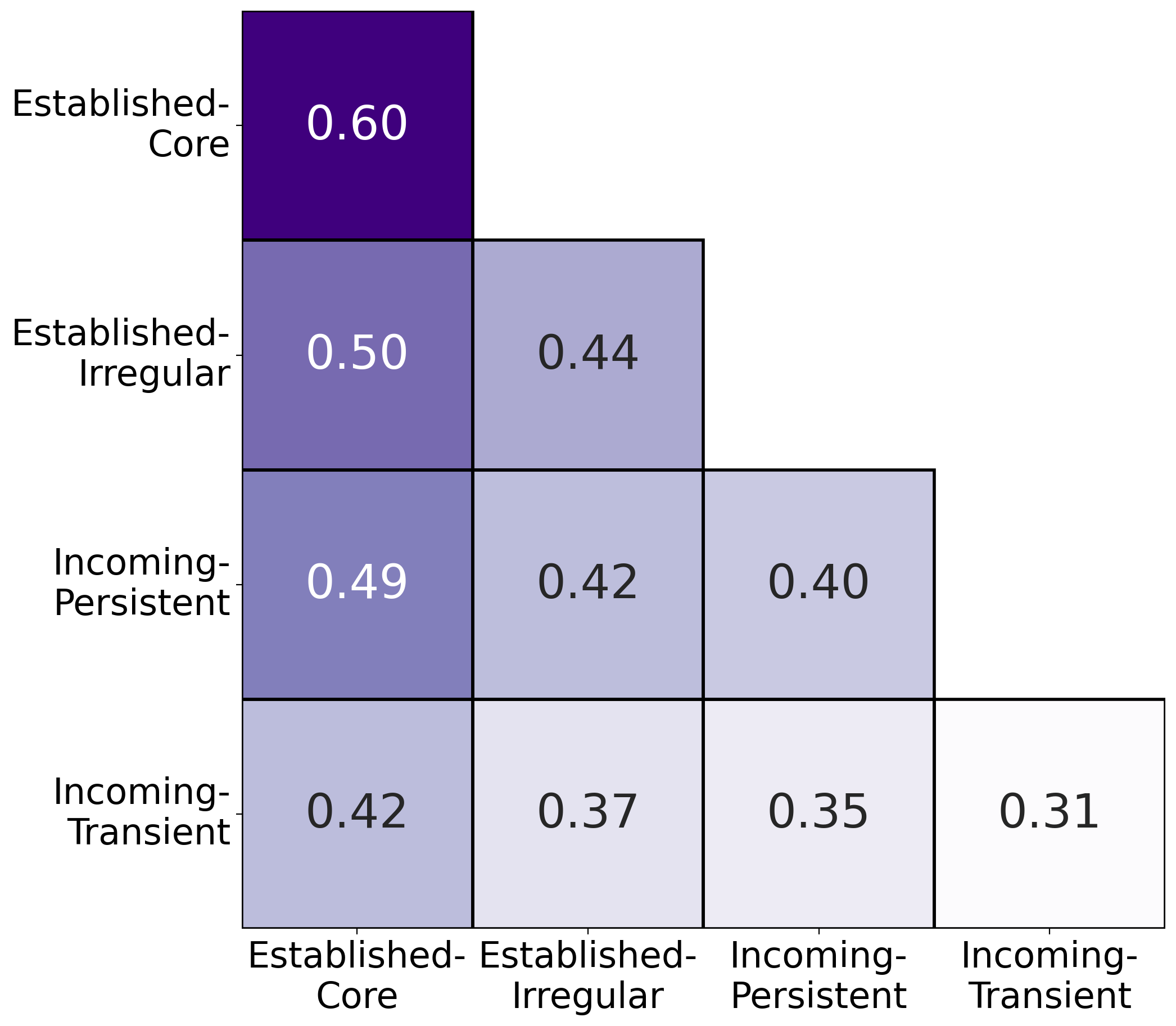}
      \textbf{B - Week 2021-28}
  \end{minipage}%
  \hspace{0.2cm}
  \begin{minipage}[b]{0.315\linewidth}
      \centering
      \includegraphics[width=\linewidth]{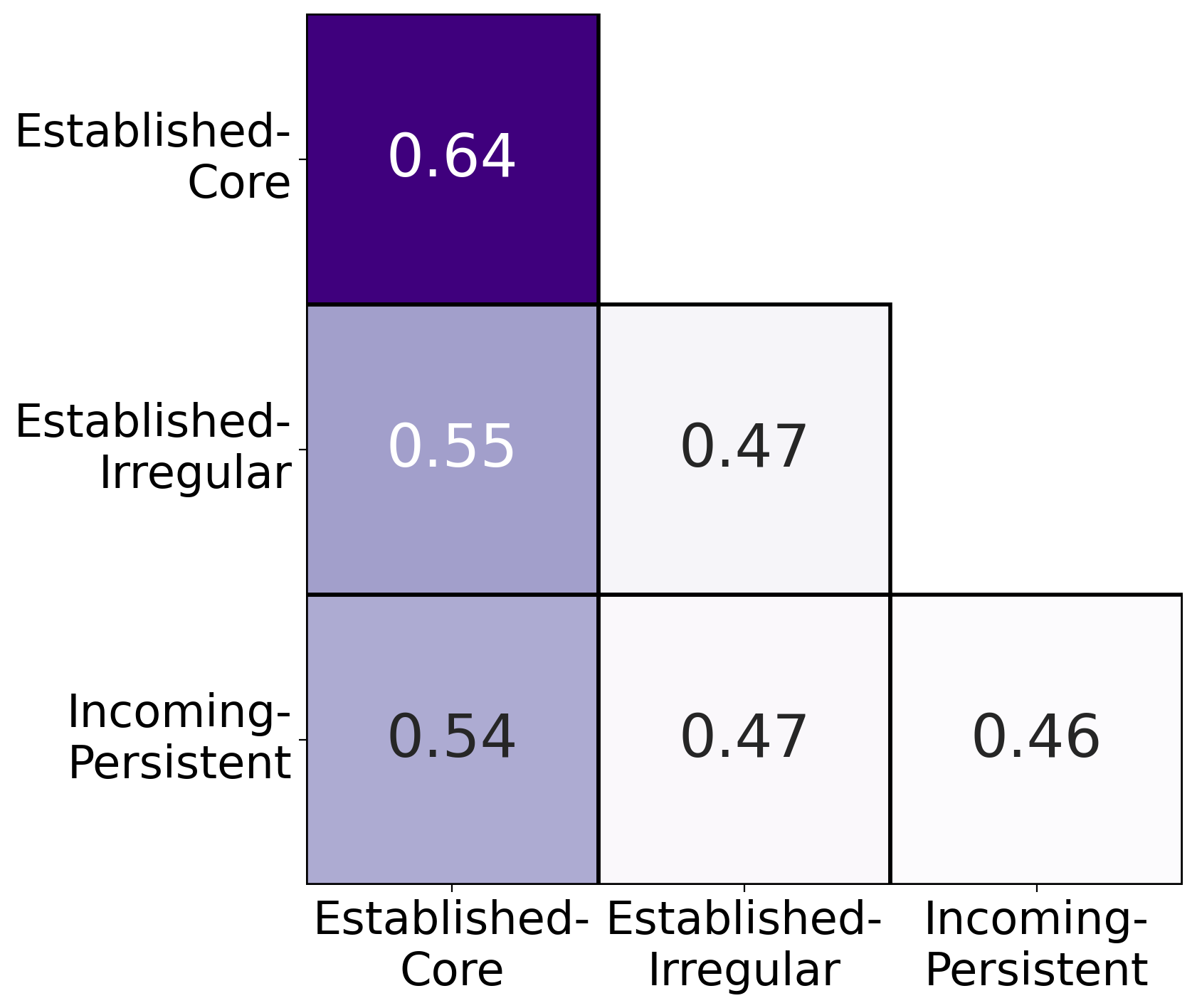}
      \textbf{C - Week 2021-29}
  \end{minipage}
  \caption{Cosine similarity between the Established-Core, Established-Irregular, Incoming-Persistent, and Incoming-Transient groups. We focused on the week before Macron's speech (A - 2021-27), the week of his speech (B - 2021-28), and the week after (C - 2021-29). The Established-Core consistently show the highest within-group similarity, which increases over time. The Incoming-Transient display the lowest within-group similarity and are the most distinct from the Established-Core. In contrast, the Incoming-Persistent are more similar to the Established-Core.}
  \Description{This figure shows three heatmaps for weeks 2021-27, 2021-28, and 2021-29. We see cosine similarity values between tweets of four groups of users: Established-Core, Established-Irregular, Incoming-Persistent, and Incoming-Transient. The within cosine similarity of the Established-Core is always the highest and it increases over weeks (.51 on week 2021-27, .60 on week 2021-28, and .64 on week 2021-29). The other groups are always more similar to the Established-Core than they are to themselves or other groups. The Incoming-Transient are the least similar to the Established-Core (.42 on week 2021-28) and also the most diverse group (within similarity of 0.31 on week 2021-28). In contrast, the Incoming-Persistent show a pattern closer to the Established-Irregular with a similarity to the Established-Core that increases between weeks 2021-28 and 2021-29.}
  \label{cosine_similarity}
\end{figure*}

To better understand the dynamics within and between different groups of users, we measured, for weeks 2021-27 (i.e., the week before Macron's speech), 2021-28 (i.e., the week of his speech), and 2021-29 (i.e., the week after his speech), the cosine similarity between tweets of different user groups (i.e., Established-Core, Established-Irregular, Incoming-Persistent, and Incoming-Transient). Figure \ref{cosine_similarity} shows our results for the different weeks.

The Established-Core consistently exhibited the highest within-group similarity, which increased steadily over the weeks (.51 in week 2021-27, .60 in week 2021-28, and .64 in week 2021-29). In week 2021-27, the Established-Irregular displayed the lowest within-group similarity (.38) and were more similar to the Established-Core than to themselves (.44). During the week of Macron's speech (2021-28), incoming users, particularly the Incoming-Transient, exhibited the lowest similarity to the Established-Core (.42) and the lowest within-group similarity (.31). The Incoming-Persistent behaved more similarly to the Established-Irregular, and we observed an increase in similarity both within groups (from .44 to .47 for the Established-Irregular and from .40 to .46 for the Incoming-Persistent) and with the Established-Core (from .50 to .55 for the Established-Irregular and from .49 to .54 for the Incoming-Persistent) between weeks 2021-28 and 2021-29.

\begin{figure}
  \centering
  \includegraphics[width=0.99\columnwidth]{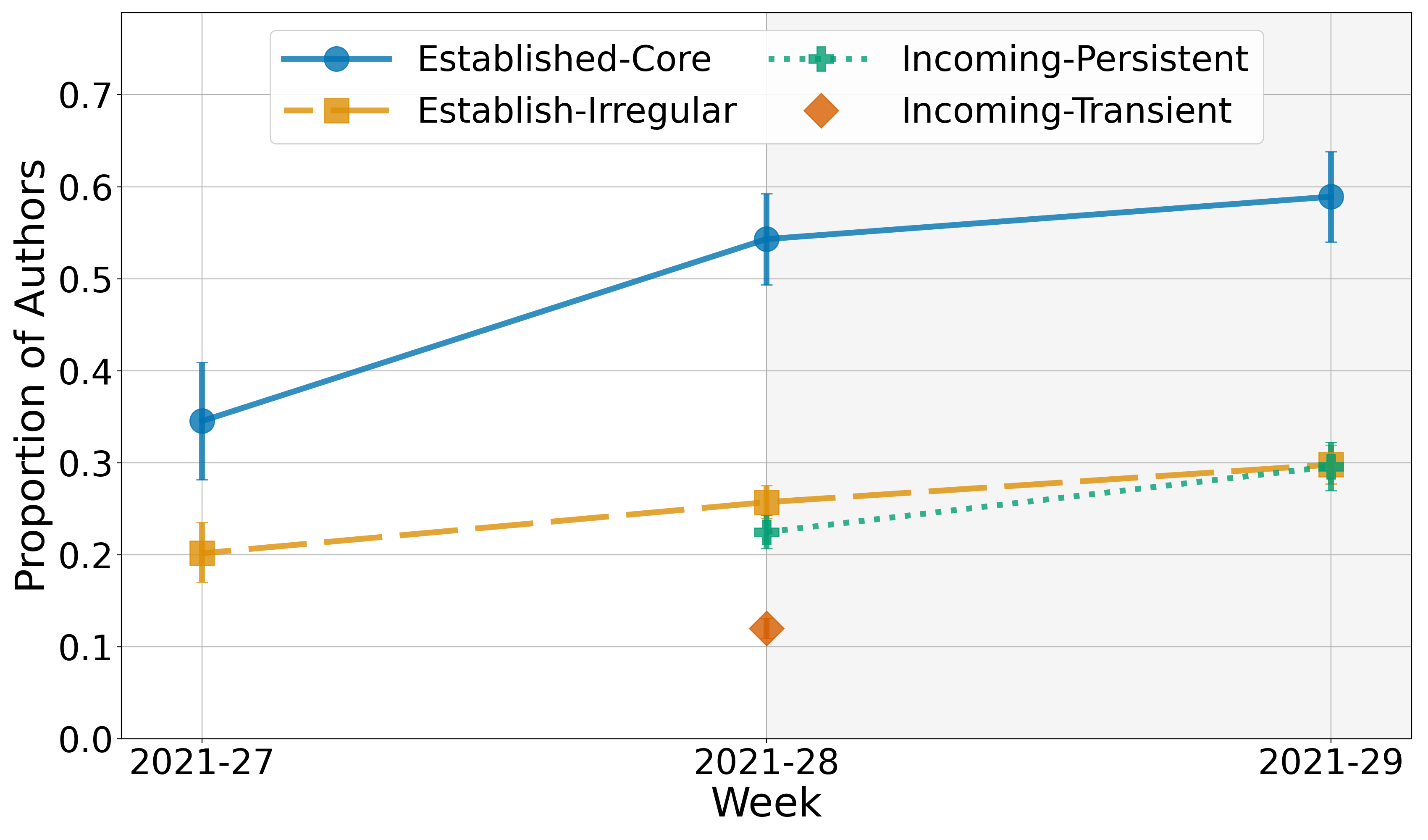}
  \caption{Proportion of outgroup label use by authors in different groups, with the period following Macron’s speech shaded grey and 95\% confidence interval error bars. The Established-Core use outgroup labels most, the Incoming-Transient least, while the Established-Irregular and Incoming-Persistent show similar patterns.}
  \Description{The Established-Core use outgroup labels the most. The Established-Irregular and Incoming-Persistent follow a similar pattern in their use of outgroup labels. The Incoming-Transient exhibit the lowest prevalence of outgroup labels.}
  \label{outgroup_groups}
\end{figure}

Additionally, we investigated the different groups' use of pronouns and outgroup labels. We found no substantial differences in pronoun usage; established users shifted from 1SG to 1PL pronouns after Macron's speech, while incoming users entered the discussion with a higher prevalence of 1PL pronouns (see Appendix for additional detail). Regarding outgroup labels (Figure \ref{outgroup_groups}), the Established-Core used them more than other groups, while the Incoming-Persistent and Established-Irregular exhibited similar trends. The Incoming-Transient used less outgroup labels.

Finally, to better understand the speech patterns that distinguish each group, we used the Fightin' Words method \cite{monroe2008fightin} to identify the most distinctive ngrams (i.e., continuous sequences of one to three words in our case) used during the week of Macron's speech (2021-28). Table \ref{fightinwords} highlights the most distinctive words for each group comparison. Z-scores indicate the importance of differences in word usage between groups, with higher absolute scores reflecting greater differences. The primary distinction lied in the focus on political, policy-related, or vaccination-related topics. The Established-Core and Incoming-Persistent were characterized by their criticism of the government and COVID-19 policies. The Established-Core emphasized opposition to the government (e.g., \#StopDictatureSanitaire \textit{[stop health dictatorship]}), \#resistance, \#GiletsJaunes \textit{[yellow vests, a previous social movement against the French government]}), while the Incoming-Persistent users focused on specific COVID-19 policies (\#PassSanitaireDeLaHonte \textit{[shame pass]}, \#NonAuPassDeLaHonte \textit{[no to the shame pass]}) and broader values (\#LiberteEgaliteFraternite \textit{[liberty, equality, fraternity]}). In contrast, the Incoming-Transient focused primarily on vaccination itself (e.g., "vaccin" \textit{[vaccine]}, "vacciner" \textit{[to vaccinate]}).

\begin{table*}[ht]
\centering
\small
\begin{tabular}{>{\raggedright\arraybackslash}p{3.2cm} >{\arraybackslash}p{6.7cm} >{\arraybackslash}p{6.7cm}}
\toprule
\textbf{Comparison} & \textbf{\centering (1)} & \textbf{\centering (2)} \\
\midrule
\textbf{Established-Core (1) vs. Incoming-Persistent (2)} & 
\#JeNeMeVaccineraiPas (24.39) \textit{[I won't get vaccinated]}, \#GiletsJaunes (19.77) \textit{[yellow vests]}, \#resistance (19.12), \#BoycottPassSanitaire (18.64) \textit{[boycott health pass]}, \#SoutienAuxSoignants (16.16) \textit{[support for healthcare workers]} & 
\#antivax (-16.47), \#NonAuPassDeLaHonte (-16.10) \textit{[no to the shame pass]}, \#LiberteEgaliteFraternite (-11.00) \textit{[liberty equality fraternity]}, medias (-10.69), \#PassSanitaireDeLaHonte (-10.49) \textit{[shame health pass]} \\
\midrule
\textbf{Established-Core (1) vs. Incoming-Transient (2)} & 
\#JeNeMeVaccineraiPas (16.84), \#resistance (16.38), \#GiletsJaunes (14.70), \#StopDictatureSanitaire (13.56) \textit{[stop health dictatorship]}, \#BoycottPassSanitaire (12.65) & 
\#antivax (-21.05), \#NousSommesDesMillions (-20.29) \textit{[we are millions]}, vaccin (-15.94) \textit{[vaccine]}, vacciner (-13.81) \textit{[to vaccinate]}, faire (-13.56) \textit{[to do]} \\
\bottomrule
\end{tabular}
\caption{Most distinctive words and Z-scores comparing different user groups. The words characteristic of group (1) in the comparison are displayed in column (1) while those for group (2) are in column (2). The comparisons suggest that the Established-Core and Incoming-Persistent frame the issue more politically, while the Incoming-Transient focus more on vaccination.}
\label{fightinwords}
\end{table*}

Vaccination remained a central issue across all groups. The hashtag \#antivax was the most distinctive term for both incoming groups, while \#JeNeMeVaccineraiPas \textit{[I won’t get vaccinated]} was prominent among Established-Core users. This suggests that while all groups expressed concerns about vaccination, the Established-Core focused more on opposing the government, the Incoming-Persistent emphasized policy critiques, and the Incoming-Transient users directly expressed their concern for vaccination.

We can interpret these results by considering incoming users as users who might be seeking to integrate into the group.\footnote{Although our results highlight only similarities and we cannot definitively determine the direction of influence (i.e., whether incoming users are becoming more similar to the Established-Core or vice versa), we will draw on previous research to guide our interpretation and provide context for understanding these findings.} There were two types of incoming users: 1) those who posted only during one week and therefore did not put so much effort into joining the group and 2) those who continued posting and therefore may be more motivated for joining the community. This pattern of behaviours is consistent with the theory of low and high identifiers~\cite{branscombe1999context}. Our findings suggested that those who kept posting and can be seen as high identifiers were also the ones most likely to conform to group norms, as evidenced by the high cosine similarity to Established-Core users, the use of 1PL pronouns, the prevalence of outgroup labels, and their most distinctive words. Having recently joined the conversation and possibly aiming to become full members of the group, these new users and high identifiers may intensify their adherence to group norms and demonstrate their commitment during this period of uncertain group membership through a higher similarity with more established members~\cite{branscombe1999context, klein2007social}. In contrast, users who posted only during one week can be considered low identifiers.

\section{Discussion and Conclusion}

In this section, we first outline the present study's limitations and propose directions for future research (\S\ref{limitations}). Then, we summarize our findings (\S\ref{conclusion}). 


\subsection{Limitations and Future Work} \label{limitations}

It is important to acknowledge that the results of this study are correlational. While we found that Macron's speech coincided with the probable crystallisation of collective identity, our measures and analyses did not establish a causal relationship. To explore this relationship further, future research could employ surveys to directly ask users about their motivations for expressing themselves on social media and their reactions to Macron’s announcement.

Additionally, the use of cosine similarity as a measure of linguistic similarity comes with limitations, as the sentence embeddings are not easily interpretable. To address this, we complemented our analysis with measures of pronoun use, outgroup references, and the most distinctive words, offering a better understanding of the linguistic parameters influencing our results.

Future studies could incorporate additional indicators, such as social network analysis to examine user connections and assess whether the ingroup becomes more cohesive over time. Sentiment analysis could also explore if members' emotions synchronize or if collective identity leads to increased positive sentiment and well-being, as suggested by previous research~\cite{branscombe1999perceiving}. Finally, to enhance the generalizability of our findings and deepen our theoretical understanding of identity emergence, it is crucial to explore this phenomenon in different contexts.

\subsection{Conclusion} \label{conclusion}

Through linguistic analysis of a year of social media discourse criticizing COVID-19 vaccines and related policies (e.g., mandatory vaccination, health pass), we gained insights into collective identity formation. We analyzed linguistic patterns—such as pronoun usage, outgroup labeling, and cosine similarity—to trace the evolution of this identity. Our findings showed that the community primarily targeted authority figures, with President Macron’s speech on mandatory vaccination and the health pass pass proving to be a pivotal event. This speech marked a shift from first-person singular to first-person plural pronouns and increased focus on vaccinated individuals as an outgroup. Additionally, many users joined the conversation during this period, and those who continued posting showed increasing linguistic similarity to established members, integrating further into the movement.

Previous research showed that institutionalized rejection based on identity can fuel major social movements~\cite{marx1998making}. In France, critics of COVID-19 vaccines may have perceived Macron's announcement as a form of institutionalized rejection, threatening core values like freedom and bodily autonomy. This perceived rejection likely strengthened collective identity among vaccine-hesitant groups. Our findings suggest that mandating vaccination may have had a counterproductive effect, reinforcing the resolve of these groups, making them more vocal and resistant, and potentially deepening societal polarization.

\begin{acks}

This publication has emanated from research conducted with the financial support of Science Foundation Ireland under Grant number 18/CRT/6049. It also received funding from the European Research Council (ERC) under the European Union’s Horizon 2020 research and innovation programme (Grant Agreement No. 802421). D. Nguyen was supported by the "Digital Society - The Informed Citizen" research programme, which is (partly) financed by the Dutch Research Council (NWO), project 410.19.007.

We also extend our gratitude to Ariana Sepahpour-Fard for manually labeling a sample of previously labeled tweets to assess annotator agreement and ensure the robustness of annotations used to train the classifier.

\end{acks}

\bibliographystyle{ACM-Reference-Format}
\bibliography{sample-base}

\appendix

\begin{figure*}[!ht]
  \centering
  \hspace*{-0.8em}
  \begin{minipage}{0.36\linewidth}
      \centering
      \includegraphics[width=\linewidth]{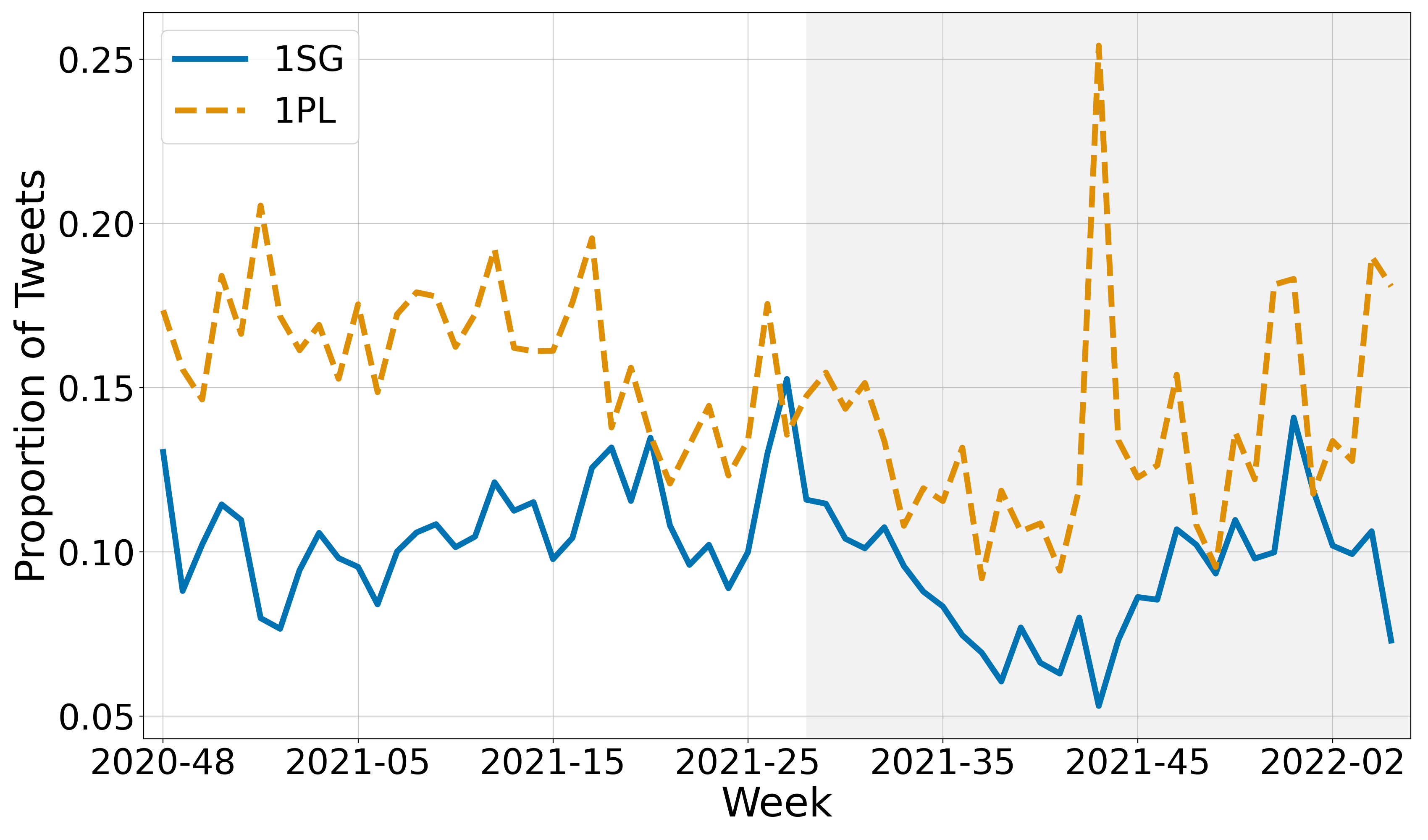}
      \textbf{A}
  \end{minipage}%
  \hspace*{-1.2em}
  \begin{minipage}{0.36\linewidth}
      \centering
      \includegraphics[width=\linewidth]{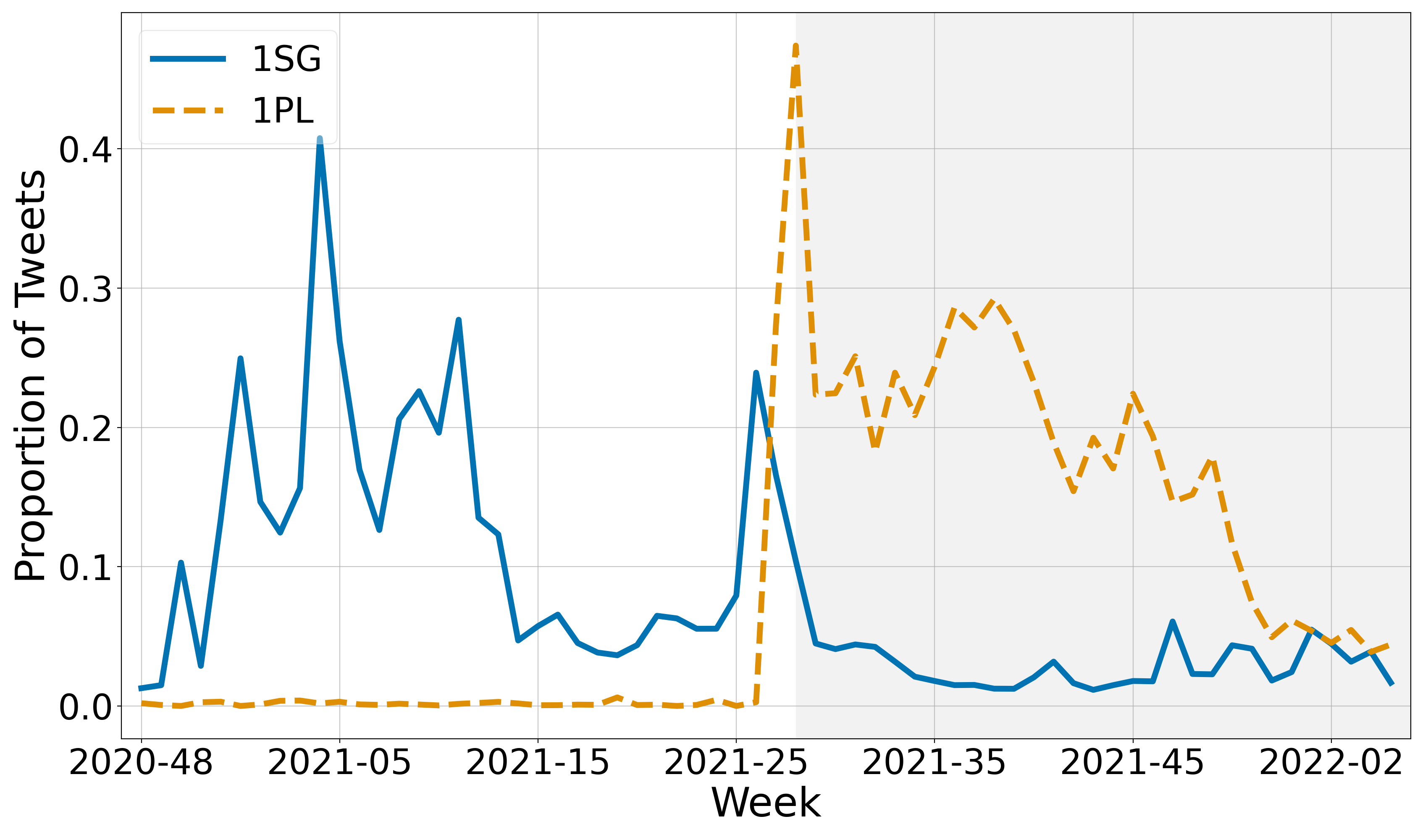}
      \textbf{B}
  \end{minipage}%
  \hspace*{-1.7em}
  \begin{minipage}{0.36\linewidth}
      \centering
      \includegraphics[width=\linewidth]{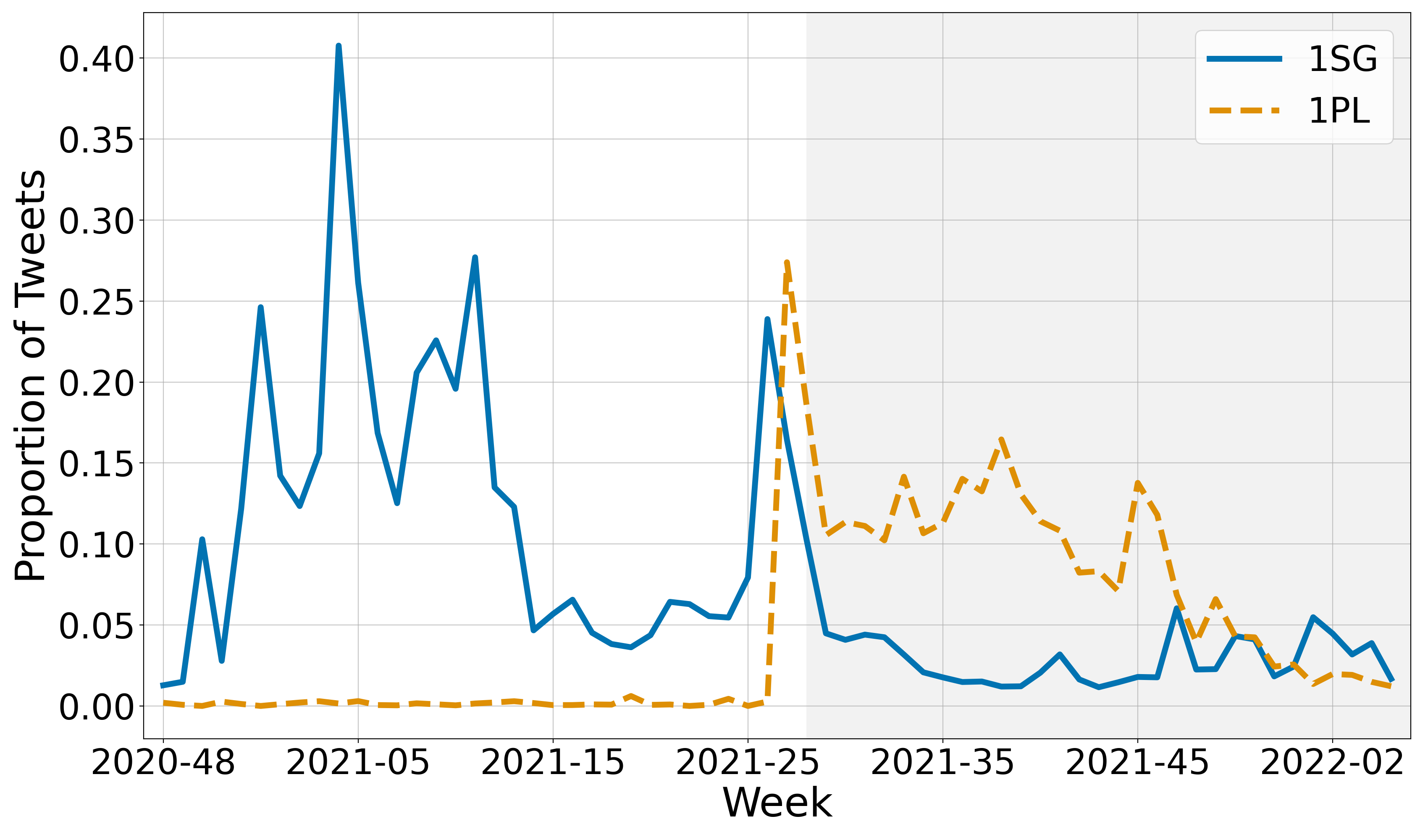}
      \textbf{C}
  \end{minipage}
  \caption{Pronoun use (A) excluding hashtags, (B) including only hashtags, and (C) including only the hashtags that were \textit{not} used for data collection; with the period following Macron’s speech shaded grey. The collective shift from first-person singular (1SG) pronouns to first-person plural (1PL) pronouns happens only when including hashtags in the analysis.}
  \Description{This figure shows three plots of the prevalence of first-person singular and first-person plural pronouns in tweets. The first plot (A) shows the prevalence of pronouns when removing hashtags. First-person plural pronouns are consistently more used than first-person singular pronouns. The second plot (B) shows the prevalence of pronouns when considering only hashtags. We see a shift on the week of Macron's speech, from first-person singular to first-person plural pronouns. The third plot (C) shows the prevalence of pronouns in tweets considering only hashtags but this time excluding the hashtags used for data collection. The shift from first-person singular to first-person plural pronouns is also visible here.}
  \label{robustness_pronouns}
\end{figure*}

\section{Classifier}

To prepare the data for our analyses (\S\ref{dataprep}), we built a classifier to identify tweets related to the criticism of COVID-19 vaccines and related policies such as the health pass. First, we sampled 1\% of the dataset (4019 tweets) for manual labeling. As participation fluctuates substantially depending on the period and external circumstances, we sampled tweets proportionally to the number of tweets published each week. In the sample used for manual classification, we identified 3,417 tweets relevant for our analysis. An external collaborator labeled 100 randomly selected tweets, and the inter-annotator agreement with our labels resulted in a Cohen's Kappa score of 0.686, indicating fair agreement ~\cite{fleiss1981measurement}.

We fine-tuned CamemBERT~\cite{martin-etal-2020-camembert} to automatically classify the entire dataset. The labeled data was partitioned into two subsets: 3500 tweets for training and validation (80\% of the total dataset, stratified proportionally to class observations, with 20\% used for validation) and 519 tweets for testing. The model was fine-tuned for ten epochs with a batch size of 16. During fine-tuning, we optimized hyperparameters using the validation set. The final model performance is presented in Table \ref{metricsclassifier}, showing detailed metrics. Notably, class 1 (tweets related to COVID-19 or related policies criticism) was well classified, achieving a precision, recall, and F1 score of .97, .96, and .96, respectively.

\begin{table}[!ht]
    \centering
    \small
    \begin{tabular}{lcccc} 
        \hline
        & Class 0 & Class 1 & Weighted Average & Macro Average \\
        \hline
        Precision & 0.783133 & 0.970183 & 0.942072 & 0.876658 \\
        Recall & 0.833333 & 0.959184 & 0.940270 & 0.896259 \\
        F1 Score & 0.807453 & 0.964652 & 0.941027 & 0.886053 \\
        \hline
        Accuracy & \multicolumn{4}{c}{0.940270} \\
        \hline
    \end{tabular}
    \caption{Metrics for each class, weighted average, macro average, and accuracy}
    \label{metricsclassifier}
\end{table}

\section{Additional Analyses for Pronoun Use}

This section presents two additional analyses related to pronoun use. First, we examine the influence of hashtags in shaping pronoun usage patterns (\S\ref{hashtagpronouns}), and second, we show the analysis comparing pronoun use of different user groups (\S\ref{pronounsgroupcomparison}).

\subsection{Hashtags and Pronoun Use} \label{hashtagpronouns}

In the main analyses (\S\ref{pronounsection}), we counted personal pronouns (i.e., "I" or "we") when they appeared at the onset of a hashtag (e.g., \#NousSommesDesMillions \textit{[We are millions]} or \#JeNeMeVaccineraiPas \textit{[I won't get vaccinated]}). We extend this analysis by examining the influence of hashtags on pronoun use. Figure \ref{robustness_pronouns} shows pronoun use when (A) excluding hashtags; (B) including only hashtags, i.e., removing the body of tweets except for hashtags; and (C) including only the hashtags that we did \textit{not} use for data collection. 

Our results indicated that hashtags played a crucial role in pronoun usage, particularly in the observed shift from first-person singular (1PL) to first-person plural (1PL) pronouns. Without hashtags, there was no discernible shift, as 1PL pronouns consistently outnumbered 1SG pronouns. However, when analyzing only hashtags (B), we saw the shift: 1SG pronouns were more prevalent before Macron's speech, while 1PL pronouns dominated afterward. To ensure that the hashtags used in our data collection were not solely responsible for this shift, we excluded them (C) and found a similar pattern, reinforcing our confidence in the results of the main analyses. In summary, our findings highlighted a shift in pronoun use following Macron's speech and underscored the importance of hashtags in this phenomenon.

\subsection{Pronoun Use in Established and Incoming Users} \label{pronounsgroupcomparison}
\begin{figure*}[!ht]
  \centering
  \includegraphics[width=0.99\linewidth]{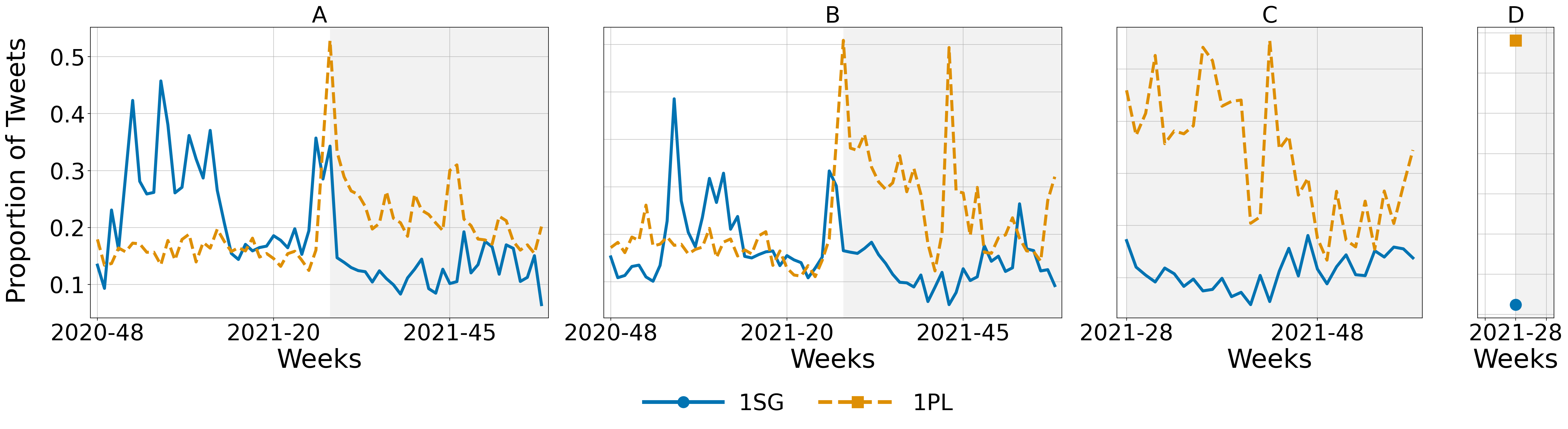}
  \caption{Pronoun use in tweets of (A) Established-Core, (B) Established-Irregular, (C) Incoming-Persistent, and (D) Incoming-Transient users; with the period following Macron’s speech shaded grey. Established users use more first-person singular (1SG) pronouns before Macron's speech while all groups use more first-person plural (1PL) pronouns from the week of his speech.}
  \label{pronoun_use_groups}
  \Description{This figure shows the use of first-person singular (1SG) and plural (1PL) pronouns in tweets of (A) Established-Core, (B) Established-Irregular, (C) Incoming-Persistent, and (D) Incoming-Transient users. Established users used more 1SG pronouns before the week of Macron's announcement while they shift to a higher use of 1PL pronouns afterwards. Incoming users use directly more 1PL pronouns than 1SG pronouns. Incoming-Persistent users show a quite stably high use of 1PL pronouns over time.}
\end{figure*}

To compare different groups of users, in particular established (i.e., users who started posting before Macron's speech) and incoming (i.e., users who started posting on the week of Macron's speech), we compared their pronoun use (in addition to the results presented in section \S\ref{incomingsection}). Figure \ref{pronoun_use_groups} shows the evolution of pronoun use for the four groups of users: (A) Established-Core, (B) Established-Irregular, (C) Incoming-Persistent, and (D) Incoming-Transient. The Established-Core and -Irregular shifted from a first-person singular (1SG) pronoun focus to a first-person plural (1PL) pronoun focus, especially visible after Macron's speech on the week 2021-28. The clearest shift from 1SG to 1PL could be observed in Established-Core users. Similarly, incoming users used more 1PL than 1SG pronouns, as they started posting on the week of Macron's speech. The high use of 1PL pronouns in Incoming-Transient users might be due to the fact that the hashtag \#NousSommesDesMillions \textit{[We are millions]} was the second most distinctive word of the group (see Table \ref{fightinwords}). Additionally, the Incoming-Persistent seemed to maintain a consistently high use of 1PL pronouns compared to established users.

\section{Outgroup Labels}

To complement the results presented in section \S\ref{outgroupsection}, we present the full list of labels and their translations in Table \ref{outgroup_table_all}.

\begin{table*}[!ht]
    \centering
    \small
    \begin{tabular}{@{}c p{5.1cm} p{5.1cm} p{5.1cm}@{}}
        \toprule
        & \textbf{Authority} & \textbf{People} & \textbf{Media} \\
        \midrule
        \multirow{2}{*}{\textbf{Neutral}} 
        & gouvernement (government), élites (elites), dirigeants (leaders), préfets (prefects), politiciens (politicians), gouvernants (governors), milliardaires (billionaires), autorités (authorities), président (president), ministres (ministers), leaders, labos (labs), laboratoires (laboratories)
        & vaccinés (vaccinated), spikés (spiked), dosés (dosed), vaxxinés (vaxxinated), vaxxés (vaxxed), provax
        & médias (media), journalistes (journalists) \\
        \cmidrule{2-4}
        \multirow{2}{*}{\textbf{Critical}} 
        & macronistes (Macron supporters), flics (cops), fdo (law enforcement), nazis, tyrans (tyrants), dictateurs (dictators), mondialistes (globalists), politicards (political hacks), ordures (scumbags), salopards (bastards), enfermistes (lockdown advocates), escrocs (crooks), maîtres (masters), pourris (rotten), collabos (collaborators with the Nazi regime), charlatans (quacks), crapules (crooks), fachos (fascists), connards (jerks), clowns, alarmistes (alarmists), fanatiques (fanatics), monstres (monsters), pourritures (scum), manipulateurs (manipulators), tarés (crazies), propagandistes (propagandists), parasites, salauds (bastards), socialistes (socialists), macronards (derogatory term for Macron supporters), menteurs (liars), khmers, ayatollahs, traîtres (traitors), fascistes (fascists), technocrates (technocrats), guignols (clowns), toubibs (docs), voyous (thugs), talibans, terroristes sanitaires (health terrorists), nantis (the wealthy)
        & moutons (sheeps), covidistes (covidists), lobotomisés (lobotomized), boomers, esclaves (slaves), psychopathes (psychopaths), vendus (sellouts), larbins (lackeys), lâches (cowards), mougeons (blend of sheeps and morons), décérébrés (brainless), crétins (morons), dociles (docile), naïfs (naive), aveugles (blind), hypocondriaques (hypochondriacs), bobos (bourgeois bohemians), marionnettes (puppets), hypnotisés (hypnotized), pleutres (cowards), toutous (lapdogs), terrorisés (terrorized), ignorants (ignorant), veaux (calves), bêtes (beasts), incultes (uncultured), obéissants (obedient), godillots (foot soldiers), sbires (henchmen), pantins (puppets), endoctrinés (indoctrinated)
        & merdias (derogatory term for media), journaleux (biased journalists), journalopes (derogatory term for journalists) \\
        \bottomrule
    \end{tabular}
    \captionof{table}{Full list of outgroup labels by category. Outgroup labels were manually categorized through two dimensions: the entity they refer to (authority, people, or media) and their stance about the entity (neutral or critical).}
    \label{outgroup_table_all}
\end{table*}

\end{document}